\documentclass[twocolumn]{article}
\usepackage{graphicx}
\usepackage{comment}
\usepackage{arxiv}
\usepackage{enumitem}

\newcommand{\todo}[1]{\textcolor{red}{#1}}
\usepackage[utf8]{inputenc} 
\usepackage[T1]{fontenc}    
\usepackage{hyperref}       
\usepackage{url}            
\usepackage{booktabs}       
\usepackage{amsfonts}       
\usepackage{nicefrac}       
\usepackage{microtype}      
\usepackage{lipsum}
\usepackage[font={footnotesize}]{caption}
\usepackage{bold-extra}
\usepackage{xcolor}
\usepackage{multirow}
\usepackage{csvsimple}
\usepackage{titlesec}
\usepackage{pbox}
\usepackage[scaled=0.85]{beramono}
\usepackage{pdfpages}
\usepackage{placeins}

\makeatletter

\newcommand{\julie}[1]{\textcolor{purple}{#1}}

\newcommand{\jannis}[1]{\textcolor{brown}{#1}}
\newcommand{\nina}[1]{\textcolor{green}{#1}}

\usepackage[page]{appendix}

\titlespacing*{\section}
{0pt}{2.5ex plus 1ex minus .2ex}{1.3ex plus .2ex}
\titlespacing{\subsubsection}{0.5pt}{\parskip}{-\parskip}
\titlespacing{\subsection}{4pt}{\parskip}{-\parskip}
\renewcommand*{\@seccntformat}[1]{\csname the#1\endcsname\hspace{2mm}}
\makeatother

\definecolor{myblue}{rgb}{0, 0, 0.7}

\usepackage{authblk}
\usepackage{setspace}
\usepackage{hyperref}
\hypersetup{colorlinks=true,
            linkcolor = myblue,
            urlcolor  = myblue,
            citecolor = myblue}

\title{\texttt{POCOVID-Net}: Automatic Detection of COVID-19 From a  New Lung Ultrasound Imaging Dataset (POCUS)}

\author[1]{Jannis Born\thanks{jborn@ethz.ch}~\textsuperscript{ ,}}
\author[2]{Gabriel Br\"andle}
\author[3]{Manuel Cossio}
\author[ ]{Marion Disdier}
\author[5]{Julie Goulet}
\author[ ]{J\'er\'emie Roulin}
\author[6]{Nina Wiedemann}

\affil[1]{Department of Biosystems Science and Engineering, ETH Zurich}
\affil[2]{Pediatric Emergencies Department, Hisrslanden Clinique des Grangettes, Geneva}
\affil[3]{Biomedical Research Institute August Pi i Sunyer, Barcelona}
\affil[5]{Physik Department and Bernstein Center for Computational Neuroscience, Technische Universit\"at M\"unchen}
\affil[6]{Department of Computer Science, ETH Zurich}

\begin{document}
\maketitle
\vspace{6mm}
\begin{abstract}
\vspace{-6mm}
With the rapid development of COVID-19 into a global pandemic, there is an ever more urgent need for cheap, fast and reliable tools that can assist physicians in diagnosing COVID-19.
Medical imaging such as CT can take a key role in complementing conventional diagnostic tools from molecular biology, and, using deep learning techniques, several automatic systems were demonstrated promising performances using CT or X-ray data.
Here, we advocate a more prominent role of point-of-care ultrasound imaging to guide COVID-19 detection.
Ultrasound is non-invasive and ubiquitous in medical facilities around the globe. \\
Our contribution is threefold. \textbf{First, we gather a lung ultrasound (POCUS) dataset} consisting of 1103 images (654 COVID-19, 277 bacterial pneumonia and 172 healthy controls), sampled from 64 videos.
This dataset was assembled from various online sources, processed specifically for deep learning models and is intended to serve as a starting point for an open-access initiative.
\\
\textbf{Second, we train a deep convolutional neural network} (\texttt{\textbf{POCOVID-Net}}) on this 3-class dataset and achieve an accuracy of 89\% and, by a majority vote, a video accuracy of 92\% .
For detecting COVID-19 in particular, the model performs with a sensitivity of 0.96, a specificity of 0.79 
and F1-score of 0.92 in a 5-fold cross validation.
\\
\textbf{Third, we provide an open-access web service} (\texttt{\textbf{POCOVIDScreen}}) that is available at:~\url{https://pocovidscreen.org}. 
The website deploys the predictive model, allowing to perform predictions on ultrasound lung images. 
In addition, it grants medical staff the option to (bulk) upload their own screenings in order to contribute to the growing public database of pathological lung ultrasound images. 
\textit{Dataset and code are available from:}~\url{https://github.com/jannisborn/covid19_pocus_ultrasound}
\textit{NOTE: This preprint is superseded by our paper in Applied Sciences (~\url{https://doi.org/10.3390/app11020672}).
}

\end{abstract}
%
%
\newpage

\vspace*{195px}
\vspace{1mm}

%
\begin{figure}[!htb]
\includegraphics[width=\columnwidth]{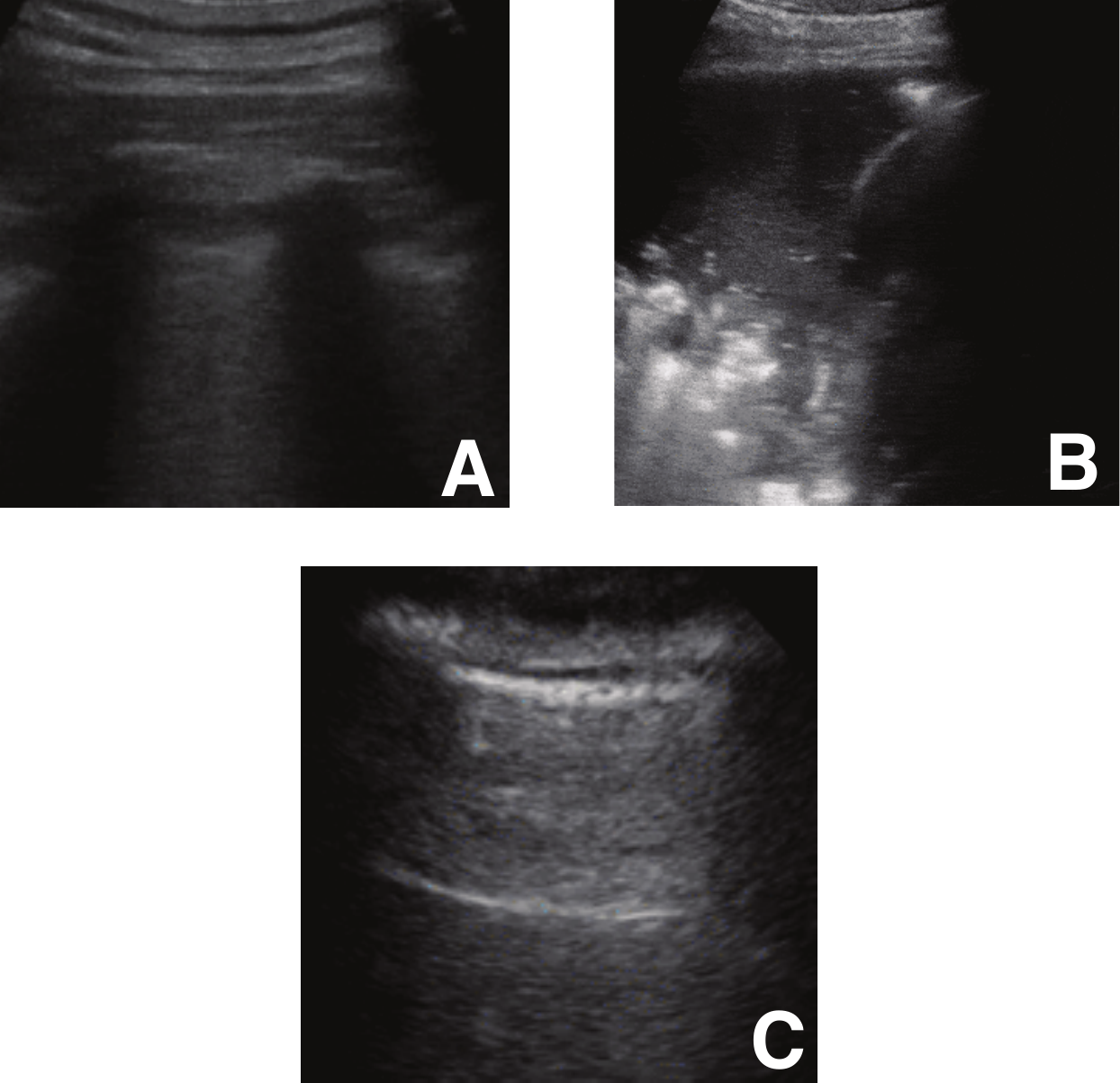}
\caption{\textbf{Example lung ultrasound images of the database.}
\textbf{A:} A typical COVID-19 infected lung, showing small subpleural consolidation and pleural irregularities.
\textbf{B:} A pneumonia infected lung, with dynamic air bronchograms surrounded by alveolar consolidation.
\textbf{C:} Healthy lung. 
The lung is normally aerated with horizontal A-lines.
All images were scraped from publicly available sources.
}
\label{fig:overview}
\end{figure}
\vspace{-4mm}
\hrule
\vspace{-2mm}
\begin{scriptsize}
\subsubsection*{Author contributions}

J.B.: Developed \& deployed model, analyzed results, wrote manuscript.\\ \setlength\parskip{2pt}
G.B.: Developed idea, approved data, reviewed manuscript.\\
M.C.: Developed idea (medical application), analyzed results, wrote manuscript.\\
M.D.: Data collection \& processing and contribution in the writing.\\
J.G.: Contributed to the background orientation, data collection, the interpretation of the statistical analysis of the model and the writing.\\
J.R.: Developed the web application and the architecture of the project.\\
N.W.: Analyzed results, deployed model, pre-processed data, wrote manuscript.\\

\end{scriptsize}

\section{Introduction}

To date, SARS-CoV-2 has infected several millions and killed hundreds of thousands  around the globe.
Due to its long incubation time, fast, accurate and reliable techniques for early disease diagnosis are key in successfully fighting the spread~\cite{li2020early}.
The standard genetic test, reverse transcription polymerase chain reaction (RT-PCR), is characterized by high reliability (at least in most countries) but a relatively long processing time (more than an hour).
Alternatively, fast serology tests are in early stages of development, and are based on antibodies that the immune system only produces in an advanced stage of the disease. 
It is of great concern, however, that several publications have reported the problem of false negatives thrown by molecular genetics and immunological tests~\cite{Imm1, RNA1, RNA2,kanne2020essentials}.
Overall, global containment efforts suffer from bottlenecks in diagnosis due to partially unreliable tests and lacking availability of the necessary testing equipment; a situation exacerbated by often asymptomatic, yet infected patients that are not properly managed due to the lack of precision around the global process~\cite{Imm1}.  

\paragraph{Biomedical imaging.}

In this context, biomedical imaging techniques have great potential to complement the conventional diagnostic techniques of COVID-19 such as molecular biology (RT-PCR) and immune (IGM/G) assays. 
Specifically, imaging can be a viable tool in the detection process by providing a fast assessment of patients in order to guide the selection of subsequent molecular and immunological tests.
Indeed, it was reported in two studies that CT scans can detect COVID-19 at higher sensitivity rate (98, respectively 88\%) compared to RT-PCR (71\% and 59\%) in cohorts of 51~\cite{fang2020sensitivity} and 1014 patients~\cite{CThigh}.
Note that sensitivity of RT-PCR varies heavily across countries, ranging from 65\%~\cite{kanne2020essentials} to 96\%~\cite{mossa2020radiology}.
While CT scans are the gold standard for pneumonia detection~\cite{bourcier2014performance}, X-ray scans are still the first line examination, despite their low specificity and sensitivity~\cite{niederman2001guidelines,weinstock2020chest}, whereas ultrasound (US) has only received growing attention in the last few years~\cite{gehmacher1995ultrasound} and achieved promising results~\cite{gazon2011agreement}.
In this contribution, we emphatically make the case for a more prominent role of the latter, based on clear evidence for the diagnostic value of lung ultrasound (US), which is provided in much detail in a comparison of CT, X-ray and US in~\autoref{sec:related}.

\paragraph{Lung point-of-care ultrasound (POCUS).}
Note first that lung ultrasound is already an established method for monitoring pneumonia and related lung diseases~\cite{gehmacher1995ultrasound,pagano2015lung,chavez2014lung}.
It has been suggested as the preferred diagnosis technique for lung infections, especially in resource limited settings such as emergency situations or low-income countries~\cite{amatya2018diagnostic} and it has started to replace X-ray as first-line examination~\cite{bourcier2014performance,gazon2011agreement,lichtenstein2004comparative,bourcier2016lung}. 
Although literature on the applicability of ultrasound for COVID-19 is still scarce, a growing body of evidence for disease-specific patterns in US has lead to advocacy for an amplified role of US in the research community~\cite{buonsenso2020covid,soldati2020there,smith2020point,sofia2020thtoracic}.

The strengths of POCUS are numerous and include its simplicity of execution, its ease of repeatability, its non-invasiveness, its execution without relocation and its ease of disinfection at the bedside.
The devices are small and portable and can be wrapped in single-use plastics to reduce the risk of contamination and promoting sterilization procedures.
Moreover, US is very cost-effective, with an estimated \$140 for an US examination compared to \$370 for chest X-ray \cite{jones2016feasibility} and \$675 – \$8600 for chest CT \cite{chestCTsource}. The low price of the device itself, starting from \$2000, 
facilitates the distribution to hospitals and primary care centers \cite{soldati2020proposal}.
The diagnostic routine can be accelerated by connecting the device to a cloud service and uploading the recordings automatically.
As a tool that is ubiquitously available even in sparsely equipped medical facilities and that can serve not only for diagnosis but also for monitoring disease evolution on a daily basis, POCUS is the ideal biomedical imaging tool in the current crisis.

The growing expectations of POCUS are perhaps best exemplified by the NIH's recently launched large-scale initiative on "Point Of Care UltraSonography (POCUS) for risk stratification of COVID-19 patients" (accessible at: \url{https://clinicaltrials.gov/ct2/show/NCT04338100}).
However, this initiative is launched without the availability of an automatic tool that can assist in COVID-19 detection and patient stratification.
Therefore, there is an evident need for an open-source framework that pools COVID-19 ultrasound scans from worldwide sources and exploits the power of deep learning to develop a system that can complement the work of physicians in a timely manner.

\paragraph{Automatic detection.}

In the last months, a myriad of preprints attempting to use machine learning for biomedical image analysis for COVID-19 has appeared, but to the best of our knowledge they exclusively focus on X-ray or CT (for reviews see~\cite{shi2020review,ulhaq2020computer,kalkreuth2020covid,pham2020artificial}) and neither a publication nor a preprint has used ultrasound data for automatic COVID-19 detection.
Here, we aim to close this gap with a first approach of training a deep learning model to detect COVID-19 on POCUS images.
It is crucial to note that medical doctors must be trained thoroughly to reliably differentiate COVID-19 from pneumonia and that the relevant patterns are hard to discern for the human eye~\cite{ng2020imaging}.
Therefore, automatic detection is highly relevant as it has been shown to reduce the time doctors invest to make a diagnosis~\cite{shan2020lung}.

\paragraph{Our contribution.}

In this work, we propose the first framework for automatized detection of COVID-19 on US images. 
Our study is in line with others demonstrating that deep learning can be a promising tool to detect COVID-19 from CT~\cite{li2020artificial} or X-ray~\cite{wang2020covid}. 
Our contributions can be summarized in the following three steps:
\begin{enumerate}[leftmargin=*,align=left]
    \item We publish the first dataset of lung POCUS recordings of COVID-19, pneumonia and healthy patients. The collected data is heterogeneous, but was pre-processed manually to remove artifacts and checked by a medical doctor for its quality.
    \item We trained a convolutional neural network (that we dub \texttt{POCOVID-Net}) on the available data and evaluated it in 5-fold cross validation.
    We report a classification accuracy of 89\% and a sensitivity to detect COVID-19 of 96\%.
    The model demonstrates the diagnostic value of the collected data and the applicability of deep learning for US images.
    \item We offer a free web service that first promotes clinical data collection by giving users the possibility to upload data, and secondly provides an interface to our trained model.
\end{enumerate}

\section{Related work}\label{sec:related}
To outline the background of the application of biomedical imaging techniques for COVID-19 diagnosis, we first compare the information content of the three main methods, and then report of previous attempts to automatize the detection.

\subsection{Biomedical imaging for COVID-19}
Three biomedical imaging sources are considered of interest for screening, diagnostics and management of COVID-19:
\paragraph{Chest X-Ray.}
Although chest X-rays were traditionally heavily used for diagnosing lung conditions, they are not well suited to detect COVID-19 at early stages~\cite{chen2020epidemiological}, for example~\cite{weinstock2020chest} found that 89\% of 493 COVID-19 patients had normal or only mildly abnormal X-ray scans. 
Instead, it is a reliable tool to evidence bilateral multifocal consolidation, partially fused into massive consolidation with small pleural effusions and "white lung”\cite{chinese2020radiological}.
However, multiple studies demonstrated the superiority of ultrasound imaging in detecting pneumonia and related lung conditions~\cite{bourcier2014performance,bourcier2016lung,reali2014can,claes2017performance}.

\paragraph{Computed tomography (CT).}
CT presents a more viable technique for early COVID-19 detection and has been the most promoted screening tool so far~\cite{bao2020covid,lee2020covid}.
Reviews report high detection rates among symptomatic individuals~\cite{bao2020covid, salehi2020coronavirus}, as CT can unveil air space consolidation, traction bronchiectasis, paving appearance and bronchovascular thickening~\cite{wang2020clinical,pan2020time}. Also, (multifocal) ground glass opacities (GGO) were observed especially frequently, often bilateral and with consolidations and prominent peripherally subpleaural distribution \cite{kanne2020chest}.
GGO are zones of increased attenuation that usually appear in several interstitial and alveolar processes with conservation of the bronchial and vascular margins \cite{franquet2011imaging}.
However, CT involves evident practical downsides such as exposing the patient to excessive radiation, high cost, the availability of sophisticated equipment (only $\sim$30,000 machines exist globally~\cite{castillo2012industry}), the need for extensive sterilization~\cite{fiala2020brief,mossa2020radiology} and patient mobilisation.

\paragraph{Ultrasound.}
Ultrasound can evidence pleural and interstitial thickening, subpleural consolidation and other physiological phenomena linked to changes in lung structure when the infection is in early stages
\cite{buonsenso2020covid}. Studies report abnormalities in bilateral B-lines (hydroaeric comet-tail artifacts arising from the pleural line), as well as identifiable lesions in the bilateral lower lobes as the main characteristics to enable COVID-19 detection~\cite{huang2020preliminary, peng2020findings}. In~\autoref{fig:overview} an example of B-lines visible in the US image of a COVID-19 patient is shown.
In a review, \cite{fiala2020brief} observed great agreement between ultrasound and CT when monitoring COVID-19 patients (especially between B-lines in ultrasound and GGOs in CT~\cite{poggiali2020can}) and concluded a high potential for ultrasound in evaluating early lung patients, especially to guide subsequent testing in triage situations.
Others reported concordance between US and chest recordings in 7/8 monitored adolescents with COVID-19, while the last patient had a normal radiography despite irregular US~\cite{sofia2020thtoracic}. 
For a review on the timeline of US findings in relation to CT see~\cite{fiala2020ultrasound}.

\subsection{Automatic detection of  
COVID-19}
While to the best of our knowledge no work has been published so far on the detection on ultrasound images, various work exists on automatic inference on CT and X-Ray scans.
\paragraph{Data collection initiatives.}
Perhaps the most significant initiative is from Cohen et. al.~\cite{cohen2020covid} who started building an open-access database of X-ray (now also CT images) that, to date, contains $\sim$150 COVID-19 images:
Using deep convolutional neural networks, many have claimed strong performances on the X-ray data of Cohen et al, ranging from
91\% up to 98\%~\cite{luz2020towards,hall2020finding,narin2020automatic,alqudahcovid,zhang2020covid,abbas2020classification,bukhari2020diagnostic}.
Besides that, the \texttt{COVID-Net} open source initiative presented the \texttt{COVIDx} dataset of chest radiography (X-ray) data~\cite{wang2020covid} that was assembled from~\cite{cohen2020covid} and other sources, resulting in 183 COVID-19 images across a total of ~13k samples. While the authors report 92~\% accuracy~\cite{wang2020covid}, others reported higher numbers with more refined models~\cite{farooq2020covid, afshar2020covid}.
Regarding CT imaging,~\cite{zhao2020covid} published a database of 275 COVID-19 CT scans and reported an accuracy of 85\%.
In light of these seemingly convincing performances, it needs to be emphasized that these databases still contain rather limited number of samples with suboptimal quality. \cite{wynants2020prediction} review 13 models proposed for the detection of COVID-19 on CT scans, and conclude that most are at high risk of bias due to qualitative and quantitative pitfalls in their data.
Since at the same time the amount of proprietary samples is rising quickly, we encourage all responsible hospitals and decision makers to contribute their data to open-access initiatives.


\renewcommand{\arraystretch}{1.3}
\begin{table*}[ht]
\resizebox{\textwidth}{!}{%
\begin{tabular}{|c|c|c|}
\hline
\textbf{Data source}   & \textbf{Data selected}                       & \textbf{Website description}                                                                                                                                                                                                                                                   \\ \hline
\textbf{Grepmed}       & 10 COVID-19, 9 pneumonia and 2 healthy videos & \begin{tabular}[c]{@{}c@{}}GrepMed is a community-sourced, searchable medical \\ image repository for referencing clinically relevant medical images\end{tabular}                                                                                                              \\ \hline
\textbf{Butterfly}     & 19 COVID-19 and 2 healthy videos                 & \begin{tabular}[c]{@{}c@{}}Butterfly is a healthtech company that launched a \\ portable US a device needing only a single probe\\ usable on the whole body that connects to a smartphone \\ (can reproduce the work of various probes such as linear and convex)\end{tabular} \\ \hline
\textbf{ThePocusAtlas} & 8 COVID-19, 3 pneumonia and 3 healthy videos & The PocusAtlas is a Collaborative Ultrasound Education Platform                                                                                                                                                                                                                \\ \hline
\end{tabular}%
}
\caption{\textbf{Data sources.} Overview of the most important data sources.}
\label{tab:data_overview}
\end{table*}

\renewcommand{\arraystretch}{1}

\vspace{-2mm}
\paragraph{Reports on propriertary data.}
Further preprints gathered their CT data independently and reported accuracy between 80\% and 90\% on 400-2700 slices~\cite{wang2020deep,xu2020deep,shi2020large} or even accuracy up to 95\% on $\sim$2000 or more slices~\cite{song2020deep,wu2020jcs,gozes2020rapid}. 
Apart from simple classification models, also segmentation technique were employed to identify infected areas and infer disease progression state to accelerate inspection time of doctors~\cite{shan2020lung, chen2020deep}.
Similarly,~\cite{li2020artificial} collected a dataset of 4,356 chest CT scans from 3,322 patients and trained a deep convolutional neural network (\texttt{COVNet}) that could differentiate COVID-19 from community-acquired pneumonia and regular scans with a ROC-AUC of 0.96 (sensitivity 90\%, specificity 96\%). 
However, to date, the data underlying all these efforts remain unavailable to the public.

\section{A lung US dataset for COVID-19 detection}

We assembled and pre-processed a dataset of a total of 64 lung POCUS video recordings, divided into  39 videos of COVID-19, 14 videos of (typical bacterial) pneumonia and 11 videos of healthy patients. Note that despite the novelty of the disease, COVID-19 images account for 60\% of our dataset.
So far, we have restricted ourselves to convex utrasound probes.
Linear and convex probes are the most standard ones in medical services, and we are focusing on the latter because more data was available.
The linear probe is a higher frequency probe and gives better resolution images, but with less tissue penetration and therefore more superficial images. The convex probe is more suitable for deep organs (abdomen, fetal ultrasound, etc.) or in obese patients. Usually, linear probes are preferred for lung ultrasound, but in practice, most medical facilities are equipped with a curved probe than can be used for everything, which explains why more convex probe images and videos were found online. However, available linear probes were collected as well, such that the model can be trained on both types of images once data availability is increased. 

\autoref{tab:data_overview} gives an overview of our sources, comprising community platforms, open medical repositories, health-tech companies and other scientific literature. Main sources of data were
\href{https://www.grepmed.com}{grepmed.com},
\href{https://thepocusatlas.com}{thepocusatlas.com},
\href{https://www.butterflynetwork.com}{butterflynetwork.com},
\href{https://radiopaedia.org}{radiopaedia.org}, while individual samples were retrieved from \href{https://everydayultrasound.com}{everydayultrasound.com}, and  \href{https://nephropocus.com}{nephropocus.com} amongst others. We provide more details on the dataset in an extensive list in Appendix~\ref{data_appendix}: First, the exact source url is given for each single single video / image file. Second, technical details are listed, comprising image resolution, the frame rate and the number of frames for each video after processing. 
Importantly, all samples of our database were observed by a medical doctor and notes on the visible patterns in each video (e.g. B-Lines or consolidations) were added.
They confirm that in all collected videos of COVID-19 and pneumonia disease-specific patterns are visible.
However, in order to train a machine learning model, a larger dataset of images instead of videos was required. 

\paragraph{Data processing.}
Since the 64 videos were taken from various sources, the format and illumination differ significantly. 
In order to generate a diverse and still sufficiently large data set, images were selected from the videos with a frame rate of 3Hz and a maximum of 30 frames per video.
This resulted in an average of 17$\pm6$ frames per video and a total of 1103 images (654 COVID-19, 277 bacterial pneumonia, 172 healthy).
To homogenize the dataset, we cropped the images with a quadratic window excluding measure bars and texts visible on the sides or top of the videos.
Five videos were manually processed with \texttt{Adobe After Effect} in order to remove measure scales and other artifacts that were overlaying the US recording.
Examples of the cropped images are shown in~\autoref{fig:overview}. We are however aware that the heterogeneity of the data is still problematic, and we are constantly searching for more data to prevent the model from over fitting on the specific properties of the available recordings.

\section{Classification with \texttt{POCOVID-Net}}

\subsection{Methods}

We propose a convolutional neural network that we name \texttt{POCOVID-Net} to tackle the present computer vision task.
First, we use the convolutional part of \texttt{VGG-16} \cite{Simonyan15}, an established deep convolutional neural network that has been demonstrated to be successful on various image types. It is followed by one hidden layer of 64 neurons with \texttt{ReLU} activation, dropout of 0.5~\cite{srivastava2014dropout} and batch normalization~\cite{ioffe2015batch}; and further by the output layer with \texttt{softmax} activation.
The model was pre-trained on \texttt{Imagenet} to extract image features such as shapes and textures.
All images are resized to $224 \times 224$ and fed through the convolutional layers of the model.
During training, only the weights of the last three layers were fine-tuned, while the other ones were frozen to the values from pre-training.
This results in a total of $2,392,963$ trainable and $12,355,008$ non-trainable parameters.
The model is trained with a the cross entropy loss function on the \texttt{softmax} outputs, and optimized with \texttt{Adam}~\cite{kingma2014adam} with an initial learning rate of $1\mathrm{e}{-4}$.

Furthermore, we use data augmentation techniques to diversify the dataset.
In explanation, the \texttt{Keras ImageDataGenerator} is used, which applies a series of random transformations on each image when generating a batch (in-place augmentation).
Here, we allow transformations of the following types: Rotations of up to 10 degrees, horizontal and vertical flips, and shifts of up to 10\% of the image height or width respectively. As such transformations can naturally occur with diverse ultrasound devices and recording parameters, augmentation adds valuable and realistic diversity that helps to prevent overfitting.

\subsection{Results}
\begin{figure}
    \centering
    \includegraphics[width=\columnwidth]{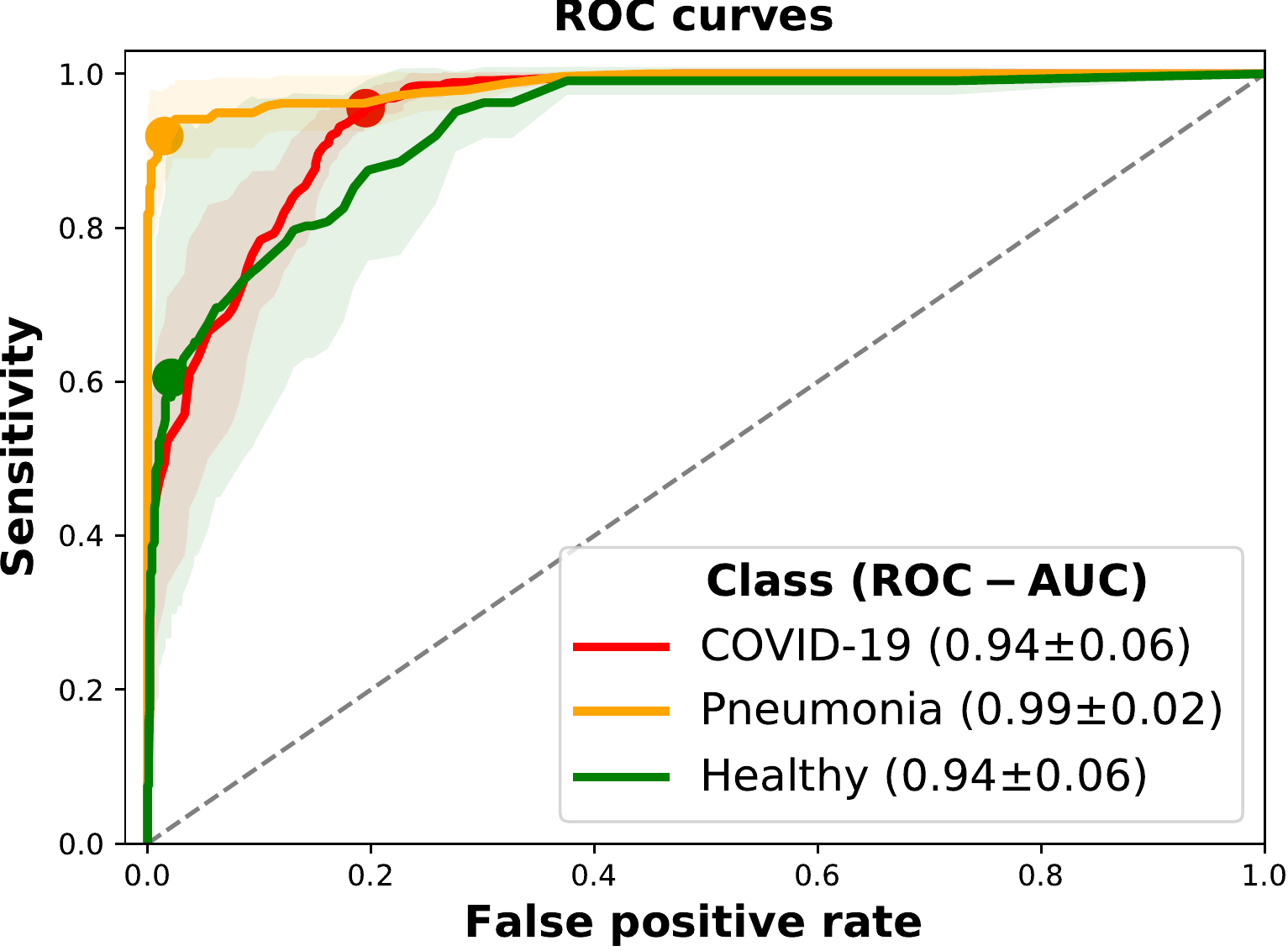}
    \caption{\textbf{Class-wise performances of \texttt{POCOVID-Net}.} Multi-class ROC curves of COVID-19 detection models from ultrasound images are depicted as averages across a 5-fold cross-validation. The shaded area shows the standard deviation of scores across folds. 
    Pneumonia is detected reliably, while the ROC-AUC for COVID-19 is 0.94 and the scores for the healthy class vary significantly. The point of highest accuracy is visualized as a coloured circle for each class.}
    \label{fig:roc}
\end{figure}
All reported results were obtained with 5-fold cross validation.
It was ensured that the frames of a single video are present within a single fold only, such that train and test data are completely disjoint recordings.\footnote{As a consequence though, the sizes of each fold vary, for example with 110 COVID-19 images in the smallest and 183 in the largest split.}

As mentioned above, the model was trained to classify frames as COVID-19, pneumonia or healthy. When splitting the data, it was assured that the number of samples per class is similar in all folds. The performance of the proposed model on the frame-wise classification task is visualized in~\autoref{fig:roc}, depicting the ROC curve for each class.

Clearly, the model learns to classify the images, with all ROC-AUC scores above or equal to 0.94. Pneumonia seems to appear very distinctive, whereas the performance on COVID-19 and regular images is lower, and varies across folds. Nevertheless, the ROC-AUC score of COVID-19 detection is 0.94.
\autoref{fig:roc} also depicts where the accuracy is maximal for each class. It can be observed that the rate of false positives at the maximal-accuracy point is larger for COVID-19 than for pneumonia and healthy patients. In a clinical setting where false positives are less problematic than false negatives, this property is desired.

\begin{figure*}[!htb]
\includegraphics[width=\textwidth]{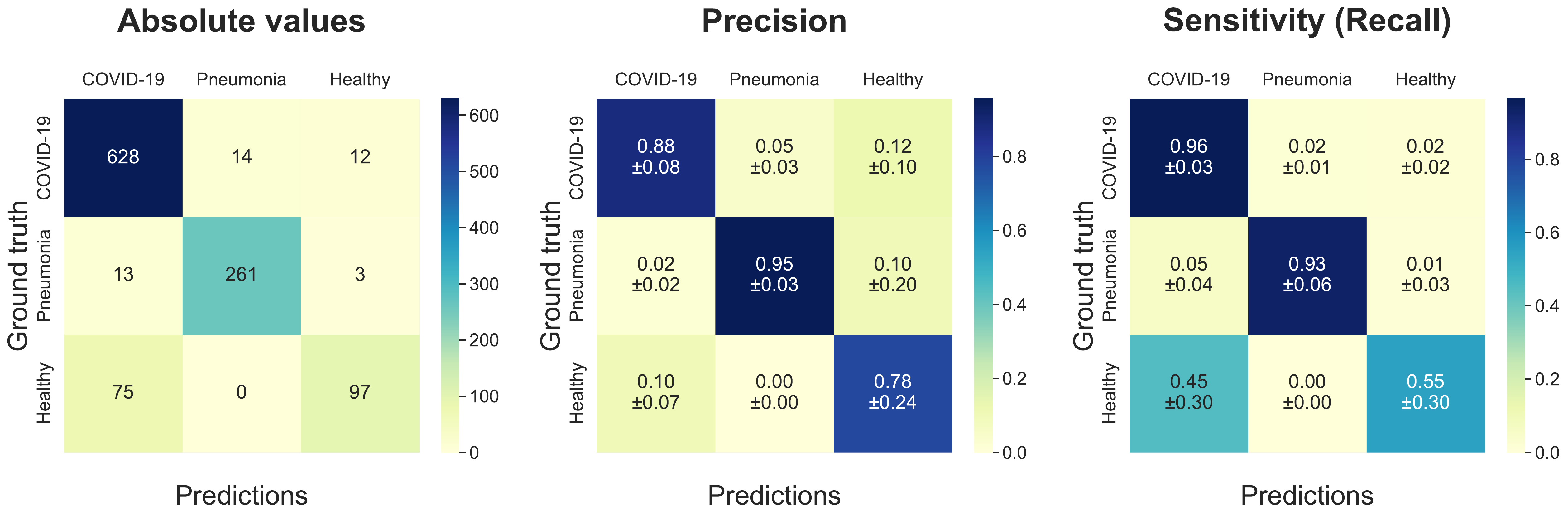}
\caption{
\textbf{Confusion matrices of POCOVID-Net on lung ultrasound images.} \textit{Left:} Absolute number of predictions.
\textit{Middle:} Relative values normalized by the number of predictions, such that precision scores can be read off the diagonal. 
\textit{Right:} Normalized by the number of ground truth members of each class, where the diagonal depicts sensitivity.
Most importantly, the sensitivity of recognizing COVID-19 is 96\%.
}
\label{fig:confusion}
\end{figure*}

Furthermore, \autoref{fig:confusion} provides more details on the predictions in the form of three confusion matrices: one with absolute values and two normalized along each axis. 

Most importantly, 628 out of 654 COVID-19 images were classified correctly, leading to a sensitivity or recall of 96\%. From the confusion matrices it becomes clear that pneumonia can be distinguished best with a sensitivity of 93\% and precision of 95\%. Note that only three frames of pneumonia were classified as healthy patients, showing - at the very least - the model's ability to recognize strong irregularities in lung images. Nevertheless, it is clear that further work is necessary to improve the false negative rate of COVID-19, as 75 images are classified as healthy lungs. We believe that a main reason for that is the low number of lung POCUS recordings of healthy subjects compared to the number of COVID-19 images. Thus, model performance might be improved significantly with more data being collected, for example in the form of collaborations with ultrasound companies or hospitals. 

To demonstrate the effectiveness of the proposed model, we compare the results to the performance of a model called COVID-Net that has recently been proposed for the classification of X-Ray images~\cite{wang2020covid}.
Training COVID-Net on our data, it achieves an accuracy of 81\% (averaged across all folds), whereas our model has 89\% accuracy. Factoring in the balanced accuracy score over the three classes (82\% vs 63\%), \texttt{POCOVID-Net} clearly outperforms \texttt{COVID-Net}.
Apart from the \texttt{COVID-Net} model, we also tested an architecture following \cite{li2020artificial} on our data. The authors employ \texttt{Res-Net}~\cite{he2016deep} instead of \texttt{VGG-16}.
In our experiments we observed that using \texttt{Res-Net} or \texttt{NasNet}~\cite{zoph2018learning} (a more recent pretrained model) resulted in significantly worse results. 

\begin{table*}[t]
  \centering
\begin{tabular}{llcccccc}
\toprule
{} &     \textbf{Class} &   \textbf{Sensitivity} &   \textbf{Specificity} &   \textbf{Precision} &   \textbf{F1-score} &   \textbf{Frames} &   \textbf{Videos/Images} \\
\midrule\\
\multirow{3}{*}{\pbox{20cm}{\textbf{POCOVID-Net}\\  \small{Acc.: 0.89} \\ Bal. Acc.: 0.82}
} &      COVID-19 &         0.96 &         0.79 &       0.88 &      0.92 &     654 &             39 \\
 &  Pneumonia &         0.93 &         0.98 &       0.95 &      0.94 &     277 &             14 \\
&    Healthy &         0.55 &         0.98 &       0.78 &      0.62 &     172 &             11 \\
\\\hline\\
\multirow{3}{*}{\pbox{20cm}{\textbf{COVID-Net}\\  \small{Acc.: 0.81} \\ Bal. Acc.: 0.63}
} &      COVID-19 &         0.98 &         0.57 &       0.77 &      0.86 &     654 &             39 \\
 &  Pneumonia &         0.89 &         0.98 &       0.95 &      0.92 &     277 &             14 \\
 &    Healthy &         0.01 &         1.00 &       0.20 &      0.01 &     172 &             11 \\
\bottomrule
\end{tabular}
\caption{\textbf{Performance comparison.} Comparison of both classification models on 5-fold cross validation for each class. Acc. abbreviates accuracy and Bal. Acc. abbreviates balanced accuracy. The proposed model outperforms COVID-Net with respect to F1-scores. In particular, COVID-Net fails to handle unbalanced classes (sensitivity for healthy patients is zero), leading to a lower specificity for COVID-19.}
\label{tab:resultsclasses}
\end{table*}

Furthermore,~\autoref{tab:resultsclasses} breaks down the comparison between \texttt{COVID-Net} and our model in more detail. An explanation for the low balanced accuracy of \texttt{COVID-Net} is given by the apparent inability of the model to deal with unbalanced data, leading to rarely any healthy-classifications. We were not able to obtain any better results when training the model on our data with the implementation provided by the authors. Although the model is able to differentiate between COVID and pneumonia to some extent (specificity of 0.98 for pneumonia), our model is superior in all scores except for sensitivity of COVID-19, resulting in a much higher false positive rate though.
In the future, it might be beneficial though to combine several different models with ensemble methods.

Finally, in an eventual clinical deployment of \texttt{POCOVID-Net} it would actually be desired to classify a video instead of single frames. We therefore summarized the frame-wise scores to determine the video class, employing two different methods: First, a majority vote of the predicted classes was taken, and secondly we average the class probabilities as predicted by the network, and then select the class with the highest average probability. Both methods achieved the same accuracy of 92\% videos that were correctly classified, and a balanced video accuracy of 84\%.

In summary, \texttt{POCOVID-Net} is able to learn frame-wise classification into COVID-19, pneumonia and healthy, where sensitivity for COVID-19 is already very high at 96\%.
It was demonstrated that the proposed architecture outperforms \texttt{COVID-Net}, and aggregating the frame-wise predictions into video classification yields a general classification performance of 92\% accuracy. Last, it was argued that further work is required to reduce the number of false positive predictions of healthy lungs as COVID-19.

\section{Web service (\texttt{POCOVIDScreen})}

The dataset and proposed model constitute a very preliminary first step toward the detection of COVID-19 from ultrasound data. 
Envisioning an interface that simplifies data sharing processes for medical practitioners thus attempting to foster collaborations with data scientists in order to serve the global need for rapid development of tools to alleviate the COVID-19 crisis, we have decided to build a web platform accessible at:~\url{https://pocovidscreen.org}.

\begin{figure}
    \centering
    \includegraphics[width=\columnwidth]{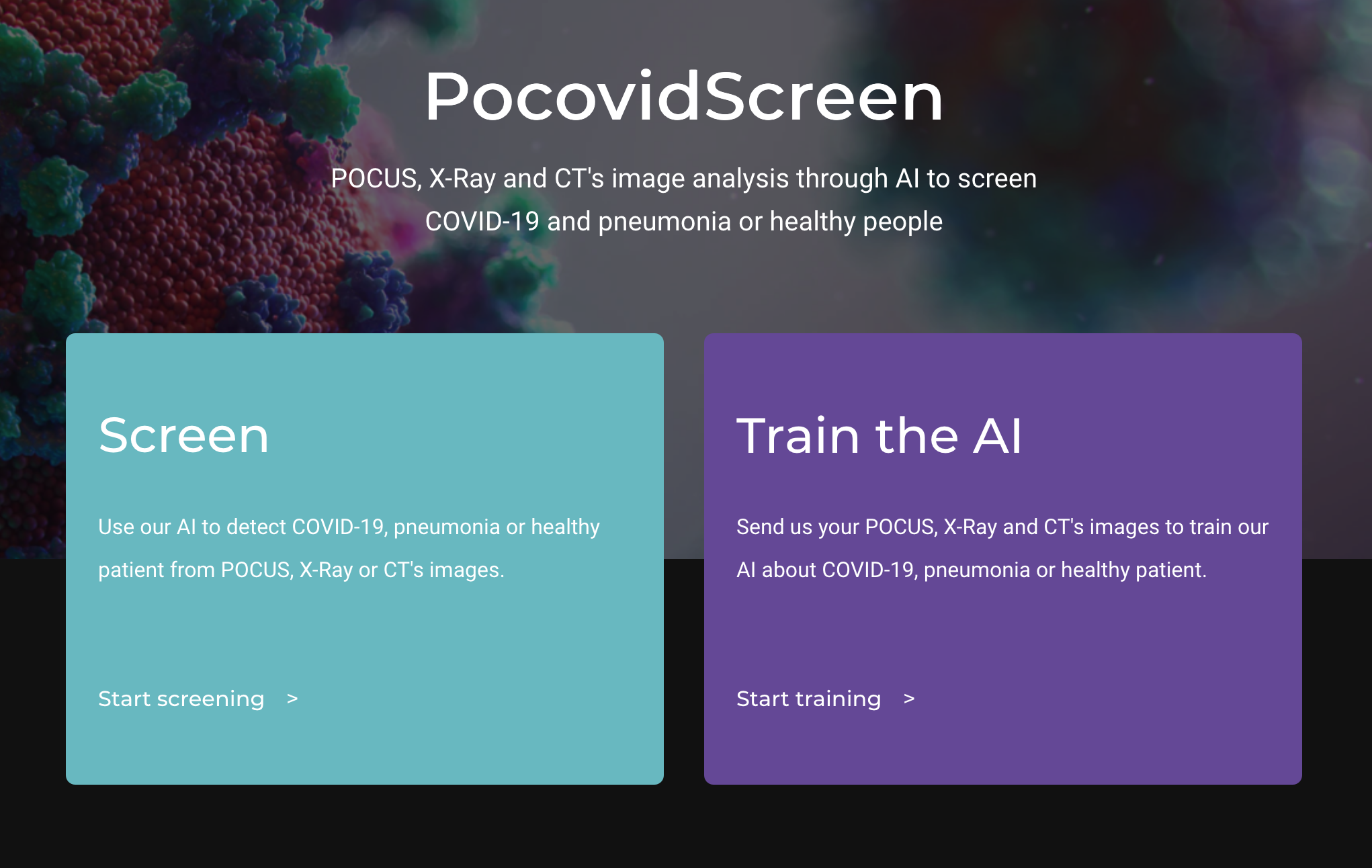}
    \caption{\textbf{Preview of our web-service \texttt{POCOVIDScreen}}. Users can upload images or videos to contribute to the dataset, or test the model \texttt{POCOVID-Net} on their own images.}
   \label{fig:preview-web}
\end{figure}

The platform (see preview in~\autoref{fig:preview-web}) is open-access and was designed for two purposes: First, users can contribute to the open-access dataset by uploading their US recordings (i.e. images or videos from COVID-19, pneumonia or healthy controls).
The collected lung ultrasound data will be continuously reviewed and approved by medical doctors, carefully processed and then integrated into our database on \texttt{GitHub}.
This strategy is chosen in order to simplify the data sharing process as much as possible.
Secondly, users can access our trained model to perform a rapid screening of their own (unlabeled) data.
Best performance is to be expected if the image is cropped to a quadratic section of the relevant part, similarly to our data.
Subsequently, the prediction is performed by evaluating all five models trained during cross validation. 
The output scores are averaged and the predicted class, in conjunction with a probability, is displayed to the user.

In short, we hope that this tool can serve as a starting point which will lead to the development of better prediction models.
Our web-service facilitates the process of data collection thus transforming the task into a community effort. Additionally it gives users the opportunity to use our model to infer the class of their own data, so far of course with a preliminary model whose outputs should not be considered of any clinical significance.

\section{Discussion}
In the current global pandemic, it is as relevant as hardly ever before that the research community pools its expertise to provide solutions in the near future.
Our contribution is the exploration of automatizing COVID-19 detection from lung ultrasound imaging in order to provide a quick assessment of the possibility of a person to be infected with COVID-19.

\paragraph{Summary}

Our first step towards this goal is to release a collection of POCUS images and videos, that were gathered and pre-processed from the referenced sources.
The videos are reliably labeled and can be split to generate a dataset of more than a thousand images.
We would like to invite researchers to contribute to our database, e.g by pointing to new publications or by making lung ultrasound recordings available.
We will constantly update this dataset on:~\url{https://github.com/jannisborn/covid19_pocus_ultrasound}.

Second, the proposed machine learning model, \texttt{POCOVID-Net}, is intended as a proof-of-concept showing the predictive power of the data, as well as the capabilities of a computer vision model to recognize certain patterns in US data as they appear in COVID-19 and pneumonia. 
Our model, based on a pre-trained convolutional network, was demonstrated a detection accuracy of 89\% and a video accuracy of 92\%.
With a high sensitivity rate of 96\% for the COVID-19 class (specificity 79\%), we provide evidence that automatic detection of COVID-19 is a promising future endeavour.
Despite these encouraging results, readers are reminded that we have presented very preliminary results herein and that we do not claim any diagnostic performance.
We also do not consider the detection on US data as a replacement for other test methods and neither as an alternative to extensive training of doctors in the usage of ultrasound for the detection task.

Last, with the presented web service we provide an interface that makes our model publicly available to researchers and hospitals around the world.
Most importantly, this interface resembles a user-friendly way to contribute lung ultrasound data to our open-access initiative.

\subsection*{Conclusion}
We believe that POCUS can provide a quick, easy and low-cost method to assess the possibility of a SARS-CoV-2 infection.
We hope to have opened a branch for automatic detection of COVID-19 from ultrasound data and envision our method as a first step toward an auxiliary tool that can assist medical doctors.
In case of a positive first-line examination, further testing is required for corroboration (e.g. CT scan or RT-PCR).
Lung ultrasound may not only take a key role in disease diagnosis, but can be utilized to monitor disease evolution through regular checks performed non-invasively and without the need of relocation.

Regarding the patient journey, we envision \texttt{POCOVID-Net} as a preliminary test for random screening and a step toward a complementary test to PCR for COVID-19 suspicions.
If a patient is suspected of COVID-19, POCUS could be used to assess the presence of lung lesion that might not have clinical repercussions and thus help detecting people at risk of lung complications which would allow to focus in a more dedicated manner on high-risk patients.
For random screening, if \texttt{POCOVID-Net} identifies the patient with COVID-19 lung symptoms, further tests such as RT-PCR, CT-scan and medical check-up should be conducted.
As POCUS devices are easily transportable, this test could take place in a primary care center where a physician or technical specialist can go to perform several screenings and avoid contamination in hospitals.
It should be noted that with one device, it is possible to perform 4 to 5 lung screenings per hour, taking into account the time needed for installation and cleaning.
 

From the machine learning perspectives several improvements to \texttt{POCOVID-Net} are possible and should be considered, given more data becomes available.
First, an evident improvement of the framework would be to perform inference directly on the videos (e.g. temporal CNNs) instead of the current frame based image analysis. 
While we did not perform this herein explicitly (due to the lack of sufficient data), we report a promising indication, namely that the video classification based on a frame-based majority vote improved the error rate by more than 25\% (from 89\% to 92\% accuracy).
Secondly, the benefit of pre-training the network on large image databases could be improved by training the model on (non lung) ultrasound samples instead of using \texttt{ImageNet}, a database of real life objects.
This pre-training may help detecting ultrasound specific patterns such as B-Lines. 
In addition, to exploit the higher availability of CT or X-ray scans, transfer learning strategies could be adopted as in \cite{zahangir2020covid_mtnet, apostolopoulos2020covid}.
Furthermore, generative models could help to complement the scarce data about COVID-19 as recently proposed in~\cite{loey2020within}.

We aim to extend the functionality of the website in the future, to further encourage the community effort of researchers, doctors and companies to build a dataset of POCUS images that can leverage the predictive power of automatic detection systems, and thereby also the value of ultrasound imaging for COVID-19 in general.
If the approach turns out to be successful, we plan to build an app as suggested in~\cite{li2020covid} that can enable medical doctors to draw inference from
their ultrasound images with unprecedented ease, convenience and speed.

\section*{Acknowledgements}
We would like to thank Jasmin Frieden for help in data processing and Moritz Gruber and J. Leo van Hemmen for feedback on the manuscript.
\section*{License}
The example images in~\autoref{fig:overview} are available via creative commons license (CC BY-NC 4.0) from:~\url{thepocusatlas.com} (access date: 17.04.2020).

\bibliographystyle{unsrt}  
\bibliography{references}  

\begin{appendices}
\section{Details on the dataset}\label{data_appendix}
We would like to acknowledge the following contributions from US videos from \href{https://radiopaedia.org}{Radiopaedia} (access date: 17.04.2020):
\begin{itemize}
    \item \href{https://radiopaedia.org/cases/pneumonia-ultrasound-1}{'Pneumonia - ultrasound' from Dr. David Carroll}
    \item \href{https://radiopaedia.org/cases/normal-anterior-lung-ultrasound-1}{'Normal anterior lung (ultrasound)' from Dr. David Carroll}
\end{itemize}
The following table presents details of the utilized videos, including source, meta information (length, size, frame rate) and comments from medical experts.

\cleardoublepage


\section*{Supplementary material}

| Filename | URL (Video Name) | Comments from web site | Comments from our MDs | Frame rate | Resolution | Length (frames) |
|---|---|---|---|---|---|---|
| Cov-Butterfly-Skip Lesion.mp4 | https://www.butterflynetwork.com/covid19/covid-19-ultrasound-gallery (Skip Lesion (44sec) -Category :Other) | | no effusion, pleural irregularities, B-lines and areas without either B-lines nor A-lines : those areas are not properly consolidatd yet but are no normal lung anymore | 30.0 | 498x542 | 1306 |
| Cov-Butterfly-COVID Lung 1.mp4 | https://www.butterflynetwork.com/covid19/covid-19-ultrasound-gallery (Confluent B Lines (27 sec)) | | B-lines | 30.0 | 496x484 | 818 |
| Cov-Atlas+(44).gif | http://www.thepocusatlas.com/covid19-1/m2ra6cg22od32eryxmdr7im3qgw714 | Confluent B lines associated with thickening and irregularity of the pleural line. A thin parapneumonic effusion can also be apreciated | consolidated | 10.0 | 145x145 | 41 |
| Cov-Atlas+(45).gif | http://www.thepocusatlas.com/covid19-1/3ytz1el5hly8cq6jigpy0y4irdmy1v | Patchy B lines associated with thickening and irregularity of the pleural line | Pleural irregularties | 10.0 | 180x180 | 21 |
| Cov-Atlas-Day+2.gif | http://www.thepocusatlas.com/covid19 (Day2) | small bilateral pleural effucion, thickened pleural line ans basal b lines | moderate effusion, B-lines, still cannot be sure about consolidation | 10.0 | 142x142 | 83 |
| Cov-Atlas-Day+4.gif | http://www.thepocusatlas.com/covid19 (Day4) | Right side on resolution, left side a more thickened pleural line+2 subpleural consolidations | Minimal to moderate effusion, pleural thickening, probably some underlying consolidation | 10.0 | 230x230 | 40 |
| Cov-clarius.gif | https://clarius.com/clinical-utility-and-technique-for-lung-ultrasound-in-covid-19-cases/ (first video on this website, video from Twitter) | B lines with variable appearance, including focal, multifocal, and confluent | B-lines | 10.0 | 313x313 | 104 |
| Cov-Butterfly-Confluent B Lines.mp4 | https://www.butterflynetwork.com/covid19/covid-19-ultrasound-gallery (Confluent B Lines (23 sec)-deleted from the website) | | Minimal effusion, pleural irregularities, lots of B-lines | 30.0 | 484x604 | 700 |
| Cov-grepmed-blines-pocus-.mp4 | https://www.grepmed.com/images/7415 | Scanning COVID-19 everyday.. but not all the B-lines confirm and not all the asymptomatic positives have no pneumonia | minimal effusion, lots of B-lines, no consolidation, fast breathing | 20.0 | 395x359 | 206 |
| Cov-grepmed2.mp4 | https://www.grepmed.com/images/7410 | Here's an example from an intensivist caring for COVID-19 patients. Note the B-lines and areas of spared pleura. | minimal effusion, lots of B-lines, no consolidation, fast breathing (if the video is real time) | 33.0 | 256x228 | 295 |
| Cov-clarius3.mp4 | https://clarius.com/clinical-utility-and-technique-for-lung-ultrasound-in-covid-19-cases/ (third video on this website, video from Twitter) | Larger consolidations with occasional air bronchograms | consolidated | 25.0 | 1080×1080 | 295 |
| Pneu-Atlas-pneumonia.gif | http://www.thepocusatlas.com/ (Classic Findings in Pneumonia) | | pleural effusion, a delimited area of consolidation in the middle and B-lines on the rigth side of the image | 10.0 | 208x208 | 59 |
| pneu-everyday.gif | https://everydayultrasound.com/blog/category/Pneumonia | In the clip, you can see liver parenchyma to the R of the screen and consolidated lung to the left. | Heavily consolidated, pleural effusion, really deep field of view.Quality of images different than most of the others | 10.0 | 372x372 | 62 |
| pneu-grep-6.gif | https://www.grepmed.com/images/5721/airbronchogram-pulmonary-pneumonia-clinical-pocus-lung | | Consolidated | 8.0 | 197x197 | 45 |
| Pneu-grep-pneumonia1.mp4 | https://www.grepmed.com/images/6903 | Here there is consolidation in the distribution of large bronchi, and subtle dynamic air bronchograms | Consolidated | 25.0 | 202x221 | 150 |
| Pneu-grep-pneumonia2.mp4 | https://www.grepmed.com/images/6876 | When lung looks like liver, it's never a good thing. Multiple air bronchograms seen | Low-fi consolidation | 30.0 | 311x321 | 418 |
| Pneu-grep-pneumonia4.mp4 | https://www.grepmed.com/images/6431 | Pneumonia with irregular step ladder or "shred sign" on Lung Ultrasound | Low-fi consolidation | 33.0 | 317x302 | 149 |
| pneu-radiopaeda.mp4 | https://radiopaedia.org/cases/pneumonia-ultrasound-1 | | Consolidated | 30.0 | 221x236 | 223 |
| Reg-Atlas-lungcurtain.gif | http://www.thepocusatlas.com/ (Lung Curtain) | | Normal lung + Interface with liver | 10.0 | 212x212 | 60 |
| Reg-Atlas.gif | http://www.thepocusatlas.com/ (Improve Lung Sliding Visualization) | | Normal lung | 10.0 | 250x250 | 59 |
| Reg-bcpocus.gif | https://www.bcpocus.ca/organscans/pneumonia/ | | I approve the classification made on the website | 10.0 | 374x374 | 61 |

| File | Source | Description | Notes | FPS | Resolution | Frames |
|---|---|---|---|---|---|---|
| Reg-Butterfly-Normal Lung A lines.mp4 | https://www.butterflynetwork.com/covid19/covid-19-ultrasound-gallery (Normal Lung A lines) | | | 30.0 | 490x536 | 733 |
| Reg-NormalLungs.mp4 | https://radiopaedia.org/cases/normal-anterior-lung-ultrasound-1 | | Normal lung | 25.0 | 148x141 | 131 |
| Reg-Youtube.mp4 | https://www.youtube.com/watch?v=VzgX9ihnmec | | 1'20 normal lung, 1'33 B lines (but non-Covid related), 1'50 à G noraml, à D B-lines, 3'44 lignes B. We used 1'20 normal lung video | 30.0 | 864x664 | 172 |
| Cov-Atlas-+(43).gif | http://www.thepocusatlas.com/covid19-1/ru6w5sgjyu1zn21jraypm2pfm0h6uf | Confluent B lines associates with a small subpleural consolidation | Pleural irregularities, lots of B-lines | 10.0 | 163x163 | 39 |
| Cov-Atlas-Day+1.gif | http://www.thepocusatlas.com/covid19 (Day1) | No lung US abnormalities | No effusion, clear consolidation in one part of the lung. We can see the limits of the pneumonia (consolidated) surrounded by healthy lung! | 10.0 | 408x408 | 33 |
| Cov-Atlas-Day+3.gif | http://www.thepocusatlas.com/covid19 (Day3) | small bilateral pleural effucion, thickened pleural line and basal B lines | moderate effusion, probable consolidation | 10.0 | 240x240 | 40 |
| Pneu-Atlas-pneumonia-AirBronch.gif | http://www.thepocusatlas.com/ (Dynamic Air Bronchograms in Pneumonia) | | consolidated | 10.0 | 341x341 | 40 |
| pneu-grep-7.gif | https://www.grepmed.com/images/1304/airbronchograms-pneumonia-sonostuff-clinical-effusion-pocus-lung | Pneumonia with Air Bronchograms, Effusion | heavily consolidated | 16.0 | 118x118 | 48 |
| Pneu-grep-pneumonia3.mp4 | https://www.grepmed.com/images/6439 | | Low-fi consolidation | 33.0 | 342x354 | 112 |
| Reg-Atlas-alines.gif | http://www.thepocusatlas.com/ (A-Lines-Normal Lung) | | | 10.0 | 165x165 | 60 |
| Reg-Grep-Alines.mp4 | https://www.grepmed.com/images/7408 | Normal Pleural Lines on Lung Ultrasound | Low quality image: poorly executed echography | 25.0 | 421x355 | 85 |
| Reg-Grep-Normal.gif | https://www.grepmed.com/images/5325 | Normal horizontal repetitive artifacts | Normal lung | 10.0 | 391x391 | 33 |
| Reg-nephropocus.mp4 | https://nephropocushome.files.wordpress.com/2019/07/a-lines-titled.gif | | Low resolution, bad quality | 23.0 | 302x256 | 73 |
| Pneu-Atlas-pneumonia2.gif | http://www.thepocusatlas.com/lung/air-bronchograms | Dynamic air bronchograms represent air bubbles moving up and down airways surrounded by alveolar consolidation | consolidated | 10.0 | 327x327 | 40 |
| Pneu-grep-shredsign-consolidation.mp4 | https://www.grepmed.com/images/7583 | Shred sign in patient with non-translobar consolidation. Phased array probe used | Consolidation, at the end of the clip, we can see the hearth in the background | 24.0 | 1080x1080 | 240 |
| Pneu-grep-bacterial-hepatization-clinical.mp4 | https://www.grepmed.com/images/7582 | "Hepatization" with fluid bronchograms and dynamic air bronchograms in patient with lobar consolidation due to bacterial pneumonia | Consolidation | 24.0 | 1080x1080 | 240 |
| Pneu-grep-pulmonary-pneumonia.mp4 | https://www.grepmed.com/images/6952 | Pneumonia with fluid and dynamic air bronchograms on Lung POCUS | Consolidation | 24.0 | 1080x1080 | 144 |
| Cov-grep-7543.mp4 | https://www.grepmed.com/images/7543 | 2 zones looked like this, the rest A-lines. Dx infectious and with Hx covid suspect. Subpleural consolidation and B-lines in posterior L upper lung field and R mid axillary | Mostly normal lung, some B lines | 24.0 | 1080x1080 | 72 |
| Cov-grep-7525.mp4 | https://www.grepmed.com/images/7525 | POCUS in primary care in Madrid - Making decisions with lung ultrasound in COVID19 patients with auscultation and normal oxygen saturation. Change attitude if there is lung involvement | B lines, pleura irregularity, pleural thickening | 24.0 | 1080x1080 | 72 |
| Cov-grep-7511.mp4 | https://www.grepmed.com/images/7511 | B-Lines and Thickened Pleura and Subpleural Consolidations | Agree with the website description | 24.0 | 1080x1080 | 192 |
| Cov-grep-7510.mp4 | https://www.grepmed.com/images/7510 | B-Lines and Pleural Thickening | Poor image, high compression artefact. But I agree with the description | 24.0 | 1080x1080 | 192 |
| Cov-grep-7507.mp4 | https://www.grepmed.com/images/7507 | Initially PCR negative with compatible symptoms, normal xray. Clinically worsened in 48h and repeat PCR positive | B lines, pleura irregularity, pleural thickening | 24.0 | 1080x1080 | 120 |
| Cov-grep-7505.mp4 | https://www.grepmed.com/images/7505 | | B lines, pleural effusion, pleural thickening | 24.0 | 1080x1080 | 144 |

| Filename | URL | Description 1 | Description 2 | FPS | Resolution | Frames |
|---|---|---|---|---|---|---|
| Cov-grep-7453.mp4 | https://www.grepmed.com/images/7453 | Normal pleura adjacent to thickened pleura with focal B lines and a small subpleural consolidation | Normal pleura adjacent to thickened pleura with focal B lines. Not sure about the consolidation but poor technique | 24.0 | 1080x1080 | 240 |
| Cov-Atlas-suspectedCovid.gif | http://www.thepocusatlas.com/covid19-1/lung-us-findings-in-hypoxic-patient-with-suspected-covid19 | Lung US in Hypoxic Patient with suspected Covid-19 | Minimal pleural thickening, few B-lines | 24.0 | 1080x1080 | 96 |
| Reg-Butterfly-Normal Lung_Example 2.mp4 | https://www.butterflynetwork.com/covid-19#lungfindings (Normal Lung example 2) | | normal | 24.0 | 1080x1080 | 96 |
| Cov-Butterfly-Coalescing B lines.mp4 | https://www.butterflynetwork.com/covid19/covid-19-ultrasound-gallery (name in video name ) | | B-lines and pleural irregularities | 24.0 | 1080x1080 | 72 |
| Cov-Butterfly-Confluent B lines_Example 2.mp4 | https://www.butterflynetwork.com/covid19/covid-19-ultrasound-gallery (name in video name ) | | B-lines | 24.0 | 1080x1080 | 96 |
| Cov-Butterfly-Consolidation with Air Bronchograms.mp4 | https://www.butterflynetwork.com/covid19/covid-19-ultrasound-gallery (name in video name ) | | B-lines | 24.0 | 1080x1080 | 504 |
| Cov-Butterfly-Consolidation.mp4 | https://www.butterflynetwork.com/covid19/covid-19-ultrasound-gallery (name in video name ) | | consolidation and pleural thickening pleural lines and pleural irregularities | 24.0 | 1080x1080 | 144 |
| Cov-Butterfly-Consolidation_Example 2.mp4 | https://www.butterflynetwork.com/covid19/covid-19-ultrasound-gallery (name in video name ) | | Consolidation | 24.0 | 1080x1080 | 120 |
| Cov-Butterfly-Consolidation_Example 3.mp4 | https://www.butterflynetwork.com/covid19/covid-19-ultrasound-gallery (name in video name ) | | consolidation and pleural thickening pleural lines and pleural irregularities | 24.0 | 1080x1080 | 96 |
| Cov-Butterfly-Consolidation_Example 5.mp4 | https://www.butterflynetwork.com/covid19/covid-19-ultrasound-gallery (name in video name ) | | With the breathing movement, the lung moves and we see in the middle of the video more healthy lung than sick lung | 24.0 | 1080x1080 | 120 |
| Cov-Butterfly-Irregular Pleura and Coalescent B-lines.mp4 | https://www.butterflynetwork.com/covid19/covid-19-ultrasound-gallery (name in video name ) | | B-lines and pleural irregularities | 24.0 | 1080x1080 | 120 |
| Cov-Butterfly-Irregular Pleura with Confluent B-lines.mp4 | https://www.butterflynetwork.com/covid19/covid-19-ultrasound-gallery (name in video name ) | | B-lines and pleural irregularities | 24.0 | 1080x1080 | 96 |
| Cov-Butterfly-Irregular Pleura with Multiple B lines.mp4 | https://www.butterflynetwork.com/covid19/covid-19-ultrasound-gallery (name in video name ) | | B-lines and pleural irregularities | 24.0 | 1080x1080 | 552 |
| Cov-Butterfly-Irregular Pleura with Trace Effusion.mp4 | https://www.butterflynetwork.com/covid19/covid-19-ultrasound-gallery (name in video name ) | | Pleural irregularities | 24.0 | 1080x1080 | 96 |
| Cov-Butterfly-Irregular Pleural Line.mp4 | https://www.butterflynetwork.com/covid19/covid-19-ultrasound-gallery (name in video name ) | | B lines and pleural irregularities | 24.0 | 1080x1080 | 96 |
| Cov-Butterfly-Irregular Pleural Line_Example 2.mp4 | https://www.butterflynetwork.com/covid19/covid-19-ultrasound-gallery (name in video name ) | | B lines and pleural irregularities | 24.0 | 1080x1080 | 96 |
| Cov-Butterfly-Patchy B lines with Sparing.mp4 | https://www.butterflynetwork.com/covid19/covid-19-ultrasound-gallery (name in video name ) | | B-lines | 24.0 | 1080x1080 | 120 |
| Cov-Butterfly-Subpleural Basal Consolidation.mp4 | https://www.butterflynetwork.com/covid19/covid-19-ultrasound-gallery (name in video name ) | | B-lines | 24.0 | 1080x1080 | 120 |
| Cov-Butterfly-Subpleural Basal Consolidation_Example 2.mp4 | https://www.butterflynetwork.com/covid19/covid-19-ultrasound-gallery (name in video name ) | | B-lines | 24.0 | 1080x1080 | 96 |



\end{appendices}

\end{document}